\documentclass[conference]{IEEEtran}
\usepackage{amsmath}
\usepackage{bm}
\usepackage{amssymb}
\usepackage{graphicx}
\usepackage{color,colortbl}
\usepackage[dvipsnames]{xcolor}
\usepackage{cleveref}
\usepackage{verbatim}
\usepackage{enumerate}
\usepackage[shortlabels]{enumitem}
\usepackage[boxed,ruled]{algorithm2e}
\usepackage{cuted, nccmath}
\usepackage{dblfloatfix}
\usepackage{float} 
\usepackage{lipsum}
\usepackage{balance}
\usepackage{cite}
\usepackage{url}
\usepackage{tabularx}
\usepackage{boldline}
\usepackage{subfigure}
\include{bibliography}

\begin{document}

\title{Dirichlet Process Approach for Radio-based Simultaneous Localization and Mapping}

\author{Jaebok Lee$^*$, Hyowon Kim$^*$, Henk Wymeersch$^\dagger$, 
Sunwoo Kim$^*$ \, \\
$^*$ Department of Electronic Engineering, Hanyang University, Seoul, South Korea \\
$\dagger$ Department of Electrical Engineering, Chalmers University of Technology, Sweden}
\thanks{This research was supported by the MSIT(Ministry of Science and ICT), Korea, under the ITRC(Information Technology Research Center) support program(IITP-2021-2017-0-01637) supervised by the IITP(Institute for Information & Communications Technology Planning & Evaluation).}
\maketitle

\begin{abstract}
    Due to 5G millimeter wave (mmWave), spatial channel parameters are becoming highly resolvable, enabling accurate vehicle localization and mapping.
    We propose a novel method of radio simultaneous localization and mapping (SLAM) with the Dirichlet process (DP).
    The DP, which can estimate the number of clusters as well as clustering, is capable of identifying the locations of reflectors by classifying signals when such 5G signals are reflected and received from various objects.
    We generate birth points using the measurements from 5G mmWave signals received by the vehicle and classify objects by clustering birth points generated over time.
    Each time we use the DP clustering method, we can map landmarks in the environment in challenging situations where false alarms exist in the measurements and change the cardinality of received signals.
    Simulation results demonstrate the performance of the proposed scheme. By comparing the results with the SLAM based on the Rao-Blackwellized probability hypothesis density filter, we confirm a slight drop in SLAM performance, but as a result, we validate that it has a significant gain in computational complexity.
\end{abstract}


\section{Introduction}\label{sec:Introduction}
5G mmWave network makes it possible to obtain high-resolution measurements in time, and angular domains using a wide bandwidth and large array antenna \cite{HWC}. 
Simultaneous localization and mapping (SLAM) incorporating mapping to detect an object reflecting or scattering 5G signals and estimating the state of the user's location and direction through the characteristics of the 5G mmWave can be performed.
However, in 5G SLAM, there is a problem of missed detection of targets due to a receiver's imperfections,  false alarms due to a channel estimation error, unknown type of landmarks.

To address these problems, several radio SLAM methods have been proposed. These related studies can be divided into two topics: radio (e.g., 5G) based SLAM and SLAM using the Dirichlet process (DP) method. In the 5G SLAM literature, several techniques based on random finite set (RFS) \cite{HWC,HWTWC,5G_tracking}are proposed.
The RFS method has the advantage of dealing with clutter, cardinality of objects that change with time, and data association, but this requires a huge computational cost.
On the other hand, SLAM using message passing (MP) \cite{joint_5GSLAM,MeyerMTT,ErikBP,vp_5G} has a balance between performance and computational complexity, but it is challenging to deal with the number of clutter and unknown objects.
The clustering method-based SLAM techniques for diffuse multipath were also studied in \cite{5G_clustering, Diffuse_multipath}. Among the clustering techniques, DP, which is a non-parametric clustering technique in which the number of clusters is not fixed, is suitable for SLAM scenarios, and various SLAM and tracking techniques using it have been studied: \cite{HDP_tracking,HDP_tracking2,DPMM_tracking} considered tracking using DP, while \cite{posegraph_SLAM,HDP_SLAM,arxiv_DPSLAM} introduced a DP SLAM technique that recognizes a landmark by performing data association using a vision sensor or clustering frames and pixels. To the best of our knowledge, DP SLAM has not been considered for radio SLAM applications. 

In this paper, we propose a new approach for 5G SLAM through DP to take advantage of robustness against ever-changing cardinality and to detect virtual anchors (VAs) caused by reflectors and scattering points (SPs) in the network environment with low complexity.
We perform data association by clustering birth points from VAs and SPs through DP.
The line-of-sight (LOS) signal received from the BS through the data association is classified, and the vehicle state is estimated using this extended Kalman filter (EKF).
We confirmed that SLAM is possible with similar performance as the PHD filter, while vehicle estimation is performed using only the  (classified)  LOS signal measurement received from BS.

\section{System Model}

\subsection{Vehicle State and Dynamics}
\begin{figure}[t]
\begin{centering}
\includegraphics[width=1\columnwidth]{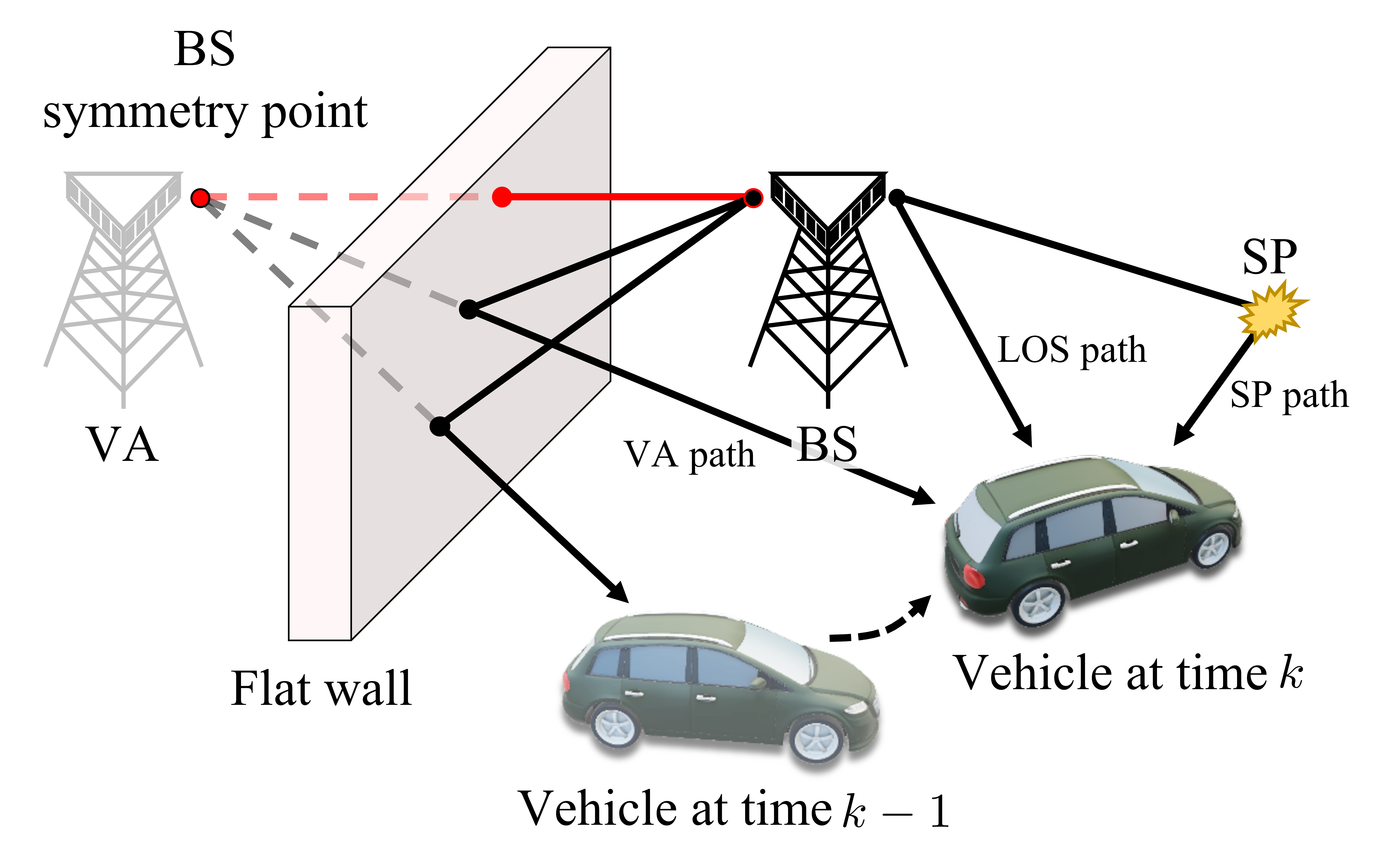}
\caption{Geographic structure of VA and SP in network environment}
\label{Fig:VA}
\end{centering}
\end{figure}
We denote a vehicle state at time $k$ by $\mathbf{s}_k =$ 
$[\mathbf{x}_{\mathbf{s},k}^\top, \alpha_{k}, \zeta_{k}, \xi_{k}, B_{k}]^\top$, where $\mathbf{x}_{\mathbf{s},k} = [x_{\mathbf{s},k},y_{\mathbf{s},k},z_{\mathbf{s},k}]^\top,$ $ \alpha_{k}, \zeta_{k}, \xi_{k},$ and $B_{k}$ are respectively 3-dimensional position, heading, translation speed, turn-rate and clock bias. 

With the known transition density $f(\mathbf{s}_k | \mathbf{s}_{k-1})$, the vehicle has dynamics  follow the motion model~\cite[Chapter 5]{ProbabilisticRobotics} over $K_{\max}$ time instant:
\begin{equation}
    \mathbf{s}_k = g(\mathbf{s}_{k-1}) + \mathbf{q}_k,
    \label{motion model}
\end{equation}
where $g(\cdot)$ is a known transition function, and $\mathbf{q}_k$ denotes a process noise, modeled as the zero-mean Gaussian distribution with the known covariance $\mathbf{Q}$.

\subsection{Propagation Environment}

The environment consists of i) a single BS, periodically transmitting the mmWave signal; ii) large surfaces, specifying VAs and reflecting the signal;
and iii) SPs, indicating small objects and scattering the signal.
A static BS location is known, denoted by $\mathbf{x}_{\text{BS}}$.
We denote VA and SP locations by $\mathbf{x}_{\text{VA}}$ and  $\mathbf{x}_{\text{SP}}$, respectively, also static. 
We regard the BS, VAs, and SPs as landmarks.
A landmark location is denoted by $\mathbf{x}_m \in \mathbb{R}^3$, and a landmark type is denoted by $m = \{\text{BS},\text{VA},\text{SP}\}$.

\subsection{Observation Model}
The mmWave signal, transmitted from the BS, is reflected by large surfaces and scattered by SPs.
At every time $k$, the vehicle receives multipath, coming from different landmarks, and observes measurements after the channel estimation routine \cite{Diffuse_multipath}.
The signal path is indexed by $i$, and the measurement of signal path $i$ is denoted by $\mathbf{z}_k^i$.
We denote a set of measurements by $\mathcal{Z}_k = \{\mathbf{z}_k^1,\ldots,\mathbf{z}_k^{I_k}\}$, while the $I_k$ is the number of paths, including LOS and non-LOS (NLOS).
Following \cite{HWTWC}, we can model the measurement $\mathbf{z}_k^i$ as
\begin{equation}
    \mathbf{z}_k^i = h(\mathbf{s}_k,\mathbf{x}^i,m)+\mathbf{r}_k^i, \label{eq:measurement}
\end{equation}
where $h(\mathbf{s}_k,\mathbf{x}^i,m) =[\tau_k^i,(\bm{\theta}_k^i)^\top,(\bm{\phi}_k^i)^\top]^\top$ and measurement noise $\mathbf{r}_k^i \sim \mathcal{N}(\mathbf{0},\mathbf{R})$ with the covariance $\mathbf{R}$. Here, $\tau_k^i$, $\bm{\theta}_k^i = [\theta_{k, \text{az}}^i, \theta_{k, \text{el}}^i]$, and $\bm{\phi}_k^i = [\phi_{k, \text{az}}^i,\phi_{k, \text{el}}^i]$ denote time of arrival~(TOA), azimuth, elevation direction of arrival~(DOA), azimuth, elevation direction of departure~(DOD) measurements. 
We denote the LOS path measurement as $\mathbf{z}_k^{\text{LOS}}$.
Due to channel estimation error, clutter may occur, included in $\mathcal{Z}_k$ with the element $\mathbf{z}_k^i$. We model clutter through $c(\mathbf{z})$, the clutter intensity which follows a Poisson point process.

\section{Overview of Dirichlet Process Clustering}
We will use DP to cluster measurements \eqref{eq:measurement} after mapping them into 3D Euclidean space. DP is a Bayesian non-parametric (BNP) model, containing an infinite number of parameters \cite{tutDP}. 

\subsection{Definition of DP}
In this section, we provide a brief overview of the process of using DP to find the probability that the data belongs to each cluster.
In the case of a finite number of clusters, the prior distribution over the clusters is defined as \cite[eq. (12)]{rasmussen}
\begin{align}
    p(l^1, \cdots ,l^D | \omega) 
    & = \frac{\Gamma(\omega)}{\Gamma(D+\omega)} \prod_{j=1}^J \frac{\Gamma(d^j+\omega/J)}{\Gamma(\omega/J)},\label{finite}
\end{align}
where $l^i$, $D$, $d^j$, $\omega$, $\Gamma(\cdot)$ are the cluster index of $i$-th data, the number of total data, the number of data assigned to $j$-th cluster, concentration parameter, and gamma function respectively. The uppercase $J$ is used to represent the total number of clusters. 
DP makes it possible for $J$ to represent a varying number of clusters, whether finite or infinite.
By using Eq. \eqref{finite}, a conditional prior for $l^i$ when all indicators other than $l^i$ are given can be easily calculated as \cite[eq. (14)]{rasmussen}
\begin{equation}
    p(l^i=j|\mathbf{l}^{-i},\omega) = \frac{d^{j}+\omega/J}{D-1+\omega},\label{-i}
\end{equation} 
where $\mathbf{l}^{-i}$ represents the set of $l$ with all indicators except the $i$.
If we let $J\to \infty$ in \eqref{-i}, then the conditional prior reaches the following equation \cite[eq. (16)]{rasmussen}
\begin{equation}
    p(l^i=j|\mathbf{l}^{-i},\omega) = \frac{d^{j}}{D-1+\omega}.
\end{equation}
Through this, the prior for a new cluster is as follows,
\begin{align}
    p(l^i=J+1|\mathbf{l}^{-i},\omega)
    &= 1 - \sum_j \frac{d^{j}}{D-1+\omega} \nonumber \\
    &= \frac{\omega}{D-1+\omega}.
\end{align}
\subsection{Data Metrics}
We also need to reflect how far apart the cluster and data are.
We quantify this using the likelihood of the Gaussian distribution through the center and covariance.
The relationship between the $y^i$ and $j$-th cluster is as follows:
we consider 
the probability of a single observation $y^i$ arising from the cluster $j$ with density $p_j(y)$ or 
from a new cluster with density $p_{0}(y)$.
We can get the conditional distribution as follows:
\begin{align}
    p(l^i=j|\mathbf{l}^{-i},\omega,y^i)
    & = p(l^i=j|\mathbf{l}^{-i},\omega)
    ~p_j(y^i) 
    ,\label{DP_ex} \\
    p(l^i=J+1|\mathbf{l}^{-i},\omega,y^i)
    & = p(l^i=J+1|\mathbf{l}^{-i},\omega) ~p_0(y^i). 
    \label{DP_new}
\end{align}
Through this, we set the cluster to which $y^i$ is assigned to the cluster with the highest assigning probability.

\section{Dirichlet Process Approach for Radio-based SLAM}

\begin{figure}[t]
\begin{centering}
\includegraphics[width=1\columnwidth]{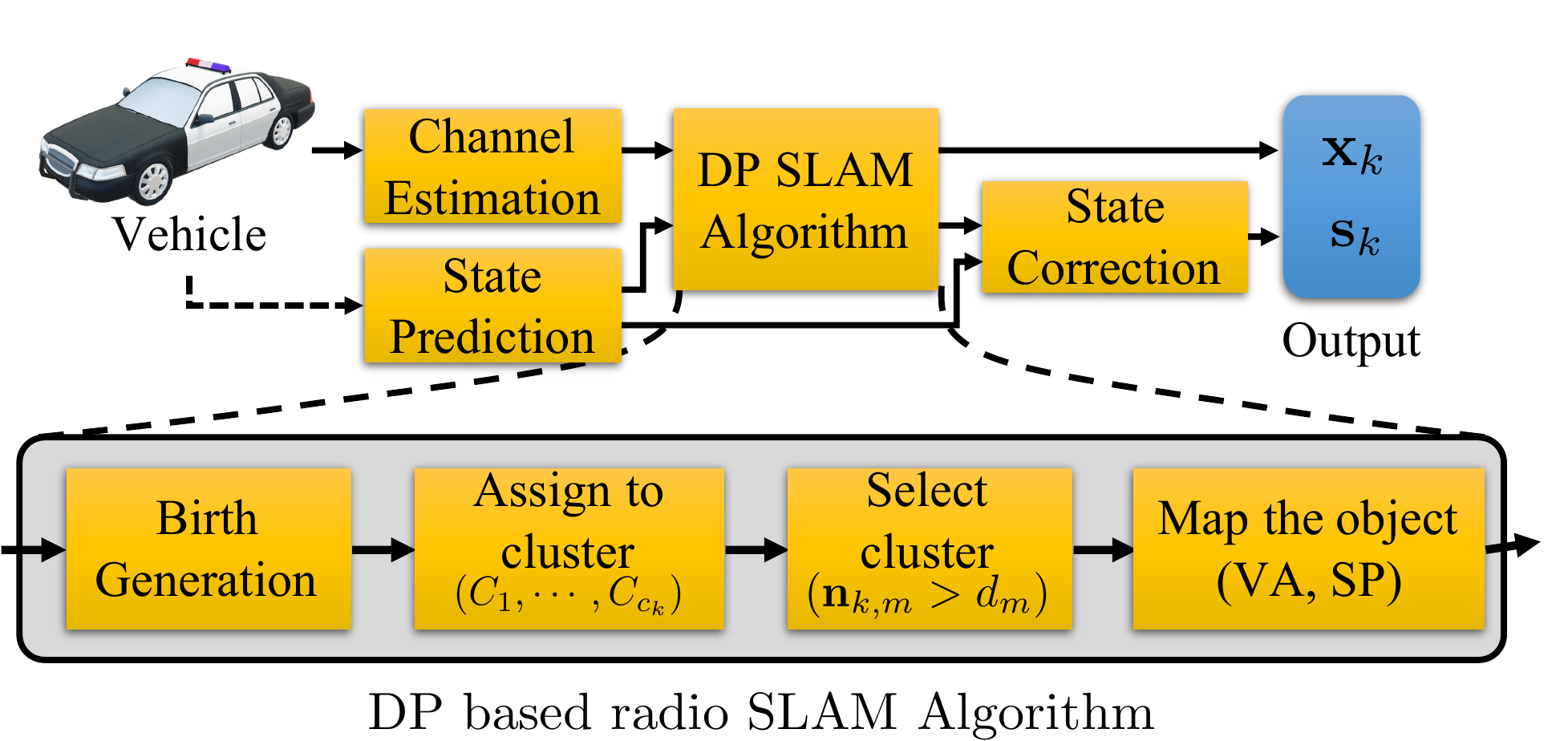}
\caption{Flow chart of the proposed DP SLAM}
\label{Fig:DP_SLAM}
\end{centering}
\end{figure}

In this section, we introduce the proposed DP-based SLAM algorithm, where clusters are modeled by Gaussian distributions. Through this algorithm, we map the objects in the environment and update the vehicle's state by classifying the LOS measurement in a situation where signals of the various path are received. The proposed algorithm consists of three steps as shown in Fig.  \ref{Fig:DP_SLAM}: vehicle state prediction, DP environment mapping, and vehicle state update.

\subsection{Vehicle Prediction}
    Given the posterior density of the vehicle state at time $k-1$, $f(\mathbf{s}_{k-1}|\mathbf{z}_{1:k-1}^\text{LOS})=\mathcal{N}(\mathbf{s}_{k-1};\widetilde{\mathbf{s}}_{k-1},\widetilde{\mathbf{V}}_{k-1})$, the vehicle density at time $k$, $f(\mathbf{s}_{k}|\mathbf{z}_{1:k-1}^\text{LOS})=\mathcal{N}(\mathbf{s}_{k};\overline{\mathbf{s}}_{k-1},\overline{\mathbf{V}}_{k-1})$, is predicted as 
\begin{align}
    f(\mathbf{s}_{k}|\mathbf{z}_{1:k-1}^\text{LOS}) = \int f(\mathbf{s}_{k}|\mathbf{s}_{k-1})f(\mathbf{s}_{k-1}|\mathbf{z}_{1:k-1}^\text{LOS})\text{d}\mathbf{s}_{k-1}.
\end{align}
    We compute $\overline{\mathbf{s}}_{k-1}$ and $\overline{\mathbf{V}}_{k-1}$ are respectively computed as
    
\begin{align}
    \overline{\mathbf{s}}_k &= g(\mathbf{s}_{k-1}),\label{vp1} \\
    \overline{\mathbf{V}}_k &= \mathbf{G}_k \widetilde{\mathbf{V}}_{k-1} \mathbf{G}^\top_k + \mathbf{Q}, \label{vp2}
\end{align}
    where $\mathbf{G}_k$ is
    a Jacobian matrix of $g(\cdot)$, computed in~\eqref{Jac_pred}, where $\mathbb{I}^{a \times b}$ and $\mathbb{O}^{a \times b}$ represents the identity matrix and zero matrix of size $a$ by $b$.
    After $\bar{\mathbf{s}}_k$ and $\bar{\mathbf{V}}_k$ are calculated, LOS measurement $\mathbf{z}_{k}^\text{LOS}$ is identified in the Section \ref{DP_clustering} and then, vehicle correction is performed using $\mathbf{z}_{k,\text{LOS}}$ in the Section \ref{vc} through EKF.

\begin{figure*}[t]
\begin{equation}
    \mathbf{G}_k = 
    \begin{bmatrix}
        \mathbb{I}^{3 \times 3} & \zeta_k/\xi_k \times \mathbf{J}_k \\
        \mathbb{O}^{4 \times 3} & \mathbf{L}
    \end{bmatrix},
    \mathbf{L} = 
        \begin{bmatrix}
            1 & 0 & 1/2 & 0 \\
            0 & 1 & 0 & 0 \\
            0 & 0 & 1 & 0 \\
            0 & 0 & 0 & 0 \\
    \end{bmatrix},
    \mathbf{J}_k = 
    \begin{bmatrix}
        \mathbf{J}^{(1)}_k & \mathbf{J}^{(2)}_k & \mathbb{O}^{4 \times 1}
    \end{bmatrix}
    ^ \top, \label{Jac_pred}
\end{equation}
\begin{equation}
    \mathbf{J}^{(1)}_k = 
    \begin{bmatrix}
    \cos(\beta_k)-\cos(\alpha_k) \\
    (\sin(\beta_k))-\sin(\alpha_k))/\zeta_k \\
    \cos(\beta_k)/2+(\sin(\alpha_k)-\sin(\beta_k))/\xi_k \\
    0
    \end{bmatrix},\nonumber
    \mathbf{J}^{(2)}_k = 
    \begin{bmatrix}
    \sin(\alpha_k + \xi_k/2)-\sin(\alpha_k) \\
    (\cos(\alpha_k)-\cos(\beta_k) )/\zeta_k \\
    \sin(\beta_k)/2-(\cos(\alpha_k)-\cos(\beta_k))/\xi_k \\
    0
    \end{bmatrix},
    \beta_k = \alpha_k+\xi_k/2.\nonumber
\end{equation} 
\hrule
\end{figure*}

\subsection{Dirichlet Process Clustering for Landmark Mapping}\label{DP_clustering}


Using the DP algorithm, we can cluster the objects' position and identify the reflectors' type. We go through the following parts: (i) initialization of clusters; (ii) birth generation from measurements, (iii) assigning to clusters. We map each type of birth point through the DP algorithm with the process Algorithm \ref{A1}. We now describe each part in detail.
\subsubsection{Initialization}
We will denote the number of clusters at time $k$ of type $m = \{\text{BS},\text{VA},\text{SP}\}$ by $J_{k,m}$. The sets containing the center, the covariance and the number of clusters of object type $m$ at time $k$ are denoted by 
$\mathcal{C}_{k,m} = \{\mathbf{c}_{k,m}^j\}_{j=1}^{J_{k,m}}  $, $\mathcal{T}_{k,m} = \{\bm{\Sigma}_{k,m}^j\}_{j=1}^{J_{k,m}}  $, and $\mathcal{D}_{k,m} = \{{d}_{k,m}^j\}_{j=1}^{J_{k,m}}$, respectively. 
We initialize the map as follows: At time $k =0$, there is no detected object and clusters but a known BS position. 
Therefore, we initialize map, 
$\mathcal{C}_{0,\text{VA}} = [\mathbf{x}_{\text{BS}}],
\mathcal{T}_{0,\text{VA}} = [\text{diag}(0.01,0.01,0.01)], \mathcal{D}_{0,\text{VA}} = [1]$ for VA map, and for SP map, $\mathcal{C}_{0,\text{SP}},
\mathcal{T}_{0,\text{SP}},
\mathcal{D}_{0,\text{SP}}$ are all empty set.

\subsubsection{Birth generation}
Before clustering, it is needed to generate birth points, where each measurement $\mathbf{z}_k$ is converted into a VA and a SP.
\begin{itemize}
    \item \emph{VA birth generation:} we generate VA birth point $\mathbf{b}_{k,\text{VA}}^i \sim \mathcal{N}(\mathbf{m}_{k,\text{VA}}^i,\mathbf{C}_{k,\text{VA}}^i)$ of measurements $\mathbf{z}_k$ with following equation,
\begin{align}
    \mathbf{m}_{k,\text{VA}}^i =& 
    \begin{bmatrix}
    \bar{x}_{s,k} + r_{k} \cos (\theta_{k, \text{az}}^i+\bar{\alpha}_{\mathbf{s},k}) \\
    \bar{y}_{s,k} + r_{k} \sin (\theta_{k, \text{az}}^i+\bar{\alpha}_{\mathbf{s},k}) \\
    \bar{z}_{s,k} + \tau_k^i \sin(\theta_{k, \text{el}}) \end{bmatrix},\label{vabirth} \\
    r_{k} =&  ~(\tau_k^i - B_k) \cos(\theta^i_{k,\text{el}}),\nonumber \\
    \mathbf{C}_{k,\text{VA}}^i =& ~(\mathbf{H}_{\mathbf{x},k}^\top \mathbf{S}_k^{-1}\mathbf{H}_{\mathbf{x},k})^{-1},\\
    \mathbf{S}_k =& ~\mathbf{H}_{\mathbf{s},k} \overline{\mathbf{V}}_k \mathbf{H}_{\mathbf{s},k}^\top,\nonumber
\end{align}
where $\mathbf{H}_{\mathbf{x},k}$, and $\mathbf{H}_{\mathbf{s},k}$ are the Jacobian matrices denoted by $\partial h/\partial \mathbf{s}_k$, and $\partial h/\partial \mathbf{x}_k$, respectively.
\item \emph{SP birth generation:} the SP birth point $\mathbf{b}_{k,\text{SP}}^i \sim \mathcal{N}(\mathbf{m}_{k,\text{SP}}^i,\mathbf{C}_{k,\text{SP}}^i)$ of the measurement can be obtained as follows,
\begin{align}
    \mathbf{m}_{k,\text{SP}}^i &= \mathbf{m}_{k,\text{VA}}^i + \frac{(\mathbf{f}_k-\mathbf{m}_{k,\text{VA}}^i)\mathbf{u}_k^\top(\mathbf{x}_{\mathbf{s},k}-\mathbf{m}_{k,\text{VA}}^i)}{(\mathbf{x}_{\mathbf{s},k}-\mathbf{m}_{k,\text{VA}}^i)\mathbf{u}_k^\top},\label{spbirth} \\
    \mathbf{u}_k &= \frac{\mathbf{x}_{\text{BS}} - \mathbf{m}_{k,\text{VA}}^i}{\Vert \mathbf{x}_{\text{BS}} - \mathbf{m}_{k,\text{VA}}^i \Vert}, \mathbf{f}_k = \frac{\mathbf{x}_{\text{BS}} + \mathbf{m}_{k,\text{VA}}^i}{2},  \nonumber \\
    \mathbf{C}_{k,\text{SP}}^i &= (\mathbf{H}_{\mathbf{x},k}^\top \mathbf{S}_k^{-1}\mathbf{H}_{\mathbf{x},k})^{-1}.
\end{align}

\end{itemize}



\begin{algorithm}[t]
\caption{DP clustering for mapping}
\SetKwInOut{Input}{input}
\SetKwInOut{Output}{output} 
\For{$m$ = \normalfont{$\{\text{VA, SP}\}$}}
    {\For{$\mathbf{z} \in \mathcal{Z}_{k,m}$}
        {{Generate birth point $\mathbf{m}_{k,m}$, according to \eqref{vabirth},  \eqref{spbirth}\;}
        \For{$j$ = \normalfont{1 to $J_{k-1,m}$}}
            {Calculate $p(l = j)$, according to \eqref{p1}\;}
        {Calculate $p(l = J_{k-1,m}+1)$, according to \eqref{p2}\;
        Find $l^*$ according to \eqref{p3}}\;
    \uIf{$l^* \le J_{k-1,m}  $ }
        {update $\bm{\Sigma}_{k,m}^{j^*}$, according to \eqref{cov}\;
        update $\mathbf{c}_{k,m}^{j^*}$, according to \eqref{center}\;}
    \Else{
    $\bm{\Sigma}_{k,m}^{j^*} =\mathbf{C}_{k,m}^i$ \;
    $\mathbf{c}_{k,m}^{j^*} =\mathbf{m}_{k,m}$ \;}
    \If{$d_{k,m}^{j^*} \ge N_{m}  $ }
    {Regard the $j^*$-th cluster as a landmark with type $m$\;}}}
\label{A1}
\end{algorithm}



\subsubsection{Clustering}
We calculate and compare each birth point's probability in an existing cluster or a new cluster to birth points map.
Using \eqref{DP_ex} and \eqref{DP_new}, the probabilities of $i$-th birth point $\mathbf{m}^{i}_{k,{m}}$ will be included in an existing clusters and a new cluster (i.e., $J_{k,m}+1$-th cluster) at time $k$ are expressed as follows, respectively,
\begin{align}
    p(l = &j \le J_{k-1,m}+1) \nonumber\\ 
    & = \mathcal{N}(\mathbf{m}^{i}_{k,{m}};\mathbf{c}^{j}_{k-1,{m}},\bm{\Sigma}_{k-1,m}^j) \frac{d^{j}_{k-1,m}}{D_{k-1}-1+\omega},
    \label{p1}  \\ 
    p(l = &J_{k-1,m}+1) \nonumber\\ 
    & = \mathcal{N}(\mathbf{m}^{i}_{k,m};{\bm{\mu}}^0,\bm{\Sigma}^0) \frac{\omega}{D_{k-1}-1+\omega},
    \label{p2}
\end{align}
where $l$ is the index of the cluster to which the $\mathbf{m}^{i}_{k,m}$ belongs.
${\bm{\mu}}^0$ is a fixed point (e.g., the origin) that represents the center of all birth points generated within the environment, $\bm{\Sigma}^0$ is a fixed large covariance of the new clusters; $\omega$ is the concentration parameter of DP, and $D_k$ means the data amount of data up to time $k$. We compare these probabilities and decide that the data belong to the cluster with the highest probability as follows,

\begin{align}
    j^* = \operatorname*{argmax}_{j\in \{1,\ldots,J_{k-1,m}+1\}} p(l=j).
    \label{p3}
\end{align}
If $j^*$ is less than or equal to $J_{k-1,m}$, which means the existing cluster is selected, the covariance and the center of the $j^*$-th cluster are updated by as follows,
\begin{align}
    \bm{\Sigma}_{k,m}^{j^*} &= (({\bm{\Sigma}_{k-1,m}^{j^*}})^{-1} +({\mathbf{C}_{k,m}^i})^{-1})^{-1},
    \label{cov}\\
    \mathbf{c}_{k,m}^{j^*} 
    &= \bm{\Sigma}_{k,m}^{j^*}
    (({\bm{\Sigma}_{k-1,m}^{j^*}})^{-1}
    \mathbf{c}_{k-1,m}^{j^*}
    +({\mathbf{C}_{k,m}^i})^{-1} \mathbf{m}^{i}_{k,m})
    .\label{center}
\end{align}
On the other hand, when a new cluster is selected, the assigned birth point's center $\mathbf{m}_{k,m}^i$ and covariance $\mathbf{C}_{k,m}^i$ are used as the center and the covariance of the cluster.
Finally, a cluster with more than object count threshold $N_{m}$ of data allocated to the cluster is recognized as a landmark.

\subsection{Vehicle Update} \label{vc}
With given measurement density, $f(\mathbf{z}_{k}^\text{LOS} |\mathbf{s}_k) = \mathcal{N}(\mathbf{z}_k^\text{LOS}; 
h(\overline{\mathbf{s}}_k, \mathbf{x}^\text{BS}, \text{BS}), \mathbf{R})$, we calculate the posterior density of the vehicle state at time $k$, $f(\mathbf{s}_{k}|\mathbf{z}_{1:k}^\text{LOS})=\mathcal{N}(\mathbf{s}_{k};\widetilde{\mathbf{s}}_{k},\widetilde{\mathbf{V}}_{k})$, as

\begin{align}
f(\mathbf{s}_{k}|\mathbf{z}_{1:k}^\text{LOS})= \eta f(\mathbf{z}_{k}^\text{LOS}|\mathbf{s}_k)f(\mathbf{s}_{k}|\mathbf{z}_{1:k-1}^\text{LOS}) 
\end{align}
where $\eta$ is a normalize term.
For the vehicle update, $\widetilde{\mathbf{s}}_{k},$ and $\widetilde{\mathbf{V}}_{k}$ are computed respectively as

\begin{align}
    \mathbf{K}_k &= \overline{\mathbf{V}}_k \mathbf{H}_k^\top(\mathbf{H}_k \overline{\mathbf{V}}_k \mathbf{H}_k^\top + \mathbf{R})^{-1},\label{vc1} \\
    \widetilde{\mathbf{s}}_k &= \overline{\mathbf{s}}_k + \mathbf{K}_k(\mathbf{z}^{\text{LOS}}_k-h(\overline{\mathbf{s}}_k,\mathbf{x}_{\text{BS}},\text{BS}))\label{vc2}, \\
    \widetilde{\mathbf{V}}_k &= (\mathbf{I}-\mathbf{K}_k \mathbf{H}_k) \overline{\mathbf{V}}_k,\label{vc3}
\end{align}
where $\mathbf{H}_k$ is the Jacobian matrix of $h( \cdot)$ at time $k$. \eqref{Jac_corr} is the expression for $\mathbf{H}_k$ when $\mathbf{x}_{\text{BS}} = [0, 0, 40]^\top$.
Finally, the vehicle state $\widetilde{\mathbf{s}}_k$ is estimated through the above series of processes.

\begin{figure*}[t]
\begin{equation}\label{Jac_corr}
\mathbf{H}_k = 
    \begin{bmatrix}
        x_{\mathbf{s},k}/\delta_k & y_{\mathbf{s},k}/\delta_k & (z_{\mathbf{s},k}-40)/\delta_k & 0 & 0 & 0 & 1  \\
        -y_{\mathbf{s},k}/\varphi_k^2 & x_{\mathbf{s},k}/\varphi_k^2 & 0 & 0 & 0 & 0 & 0 \\
        -x_{\mathbf{s},k}(z_{\mathbf{s},k}-40)/\delta_k^2\varphi_k & -y_{\mathbf{s},k}(z_{\mathbf{s},k}-40)/\delta_k^2\varphi_k & \varphi_k/\delta_k^2 &0 &0 &0 &0\\
        -y_{\mathbf{s},k}/\varphi_k^2 & x_{\mathbf{s},k}/\varphi_k^2 & 0 & -1 & 0 & 0 & 0 \\ x_{\mathbf{s},k}(z_{\mathbf{s},k}-40)/\delta_k\varphi_k & y_{\mathbf{s},k}^2(z_{\mathbf{s},k}-40)/\delta_k^2\varphi_k & -\varphi_k/\delta_k^2 &0 &0 &0 &0  \\
    \end{bmatrix},
\end{equation}
\begin{equation}
    \delta_k = \sqrt{x_{\mathbf{s},k}^2+y_{\mathbf{s},k}^2+(z_{\mathbf{s},k}-40)^2},
    ~\varphi_k = \sqrt{x_{\mathbf{s},k}^2+y_{\mathbf{s},k}^2}.
\end{equation}
\hrule
\end{figure*}

\subsection{Example of DP in Action}
\begin{figure}[t]
\centering  
\subfigure[\label{Fig:birtha}]{\includegraphics[width=0.45\linewidth]{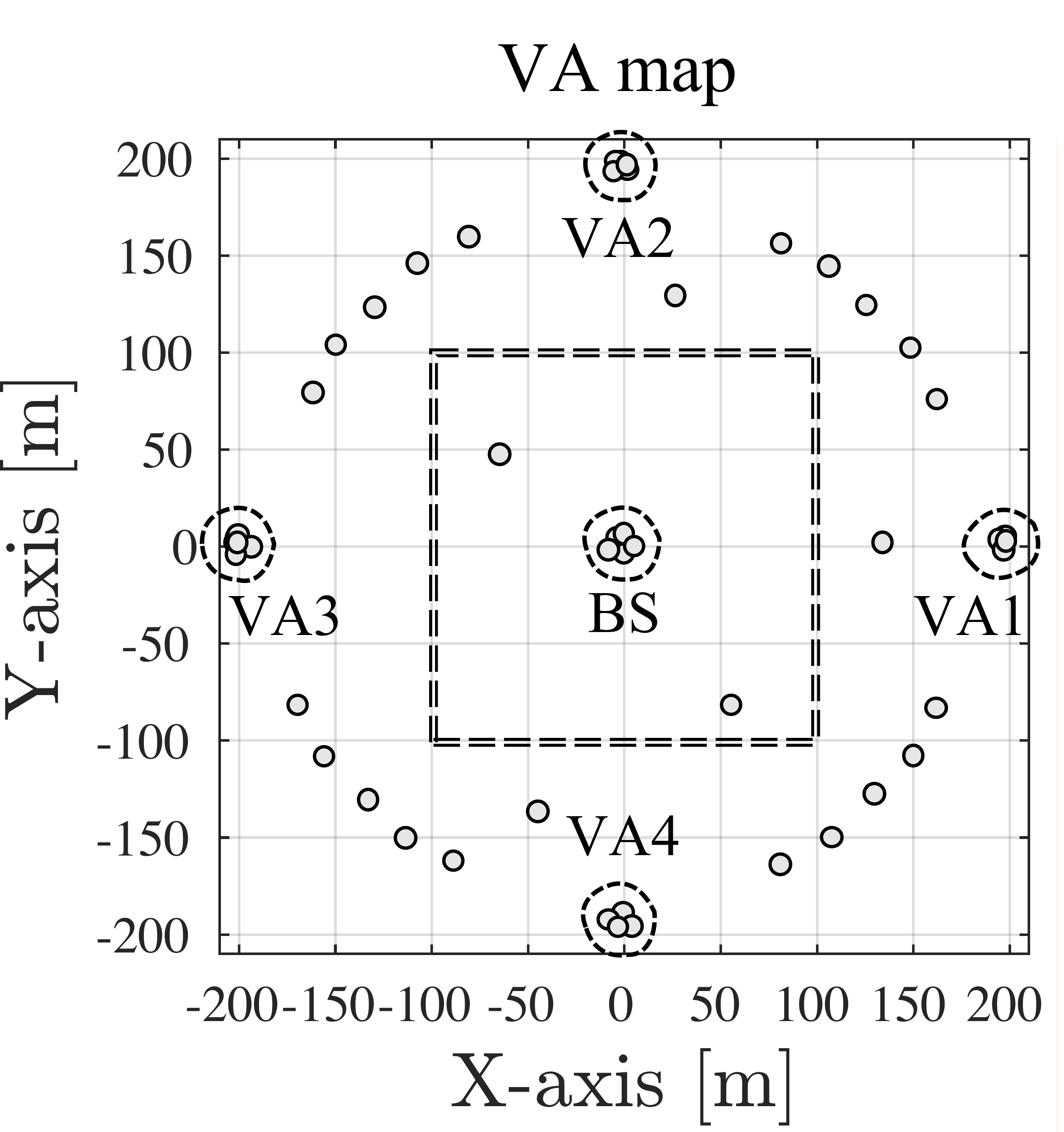}}
\subfigure[\label{Fig:birthb}]{\includegraphics[width=0.45\linewidth]{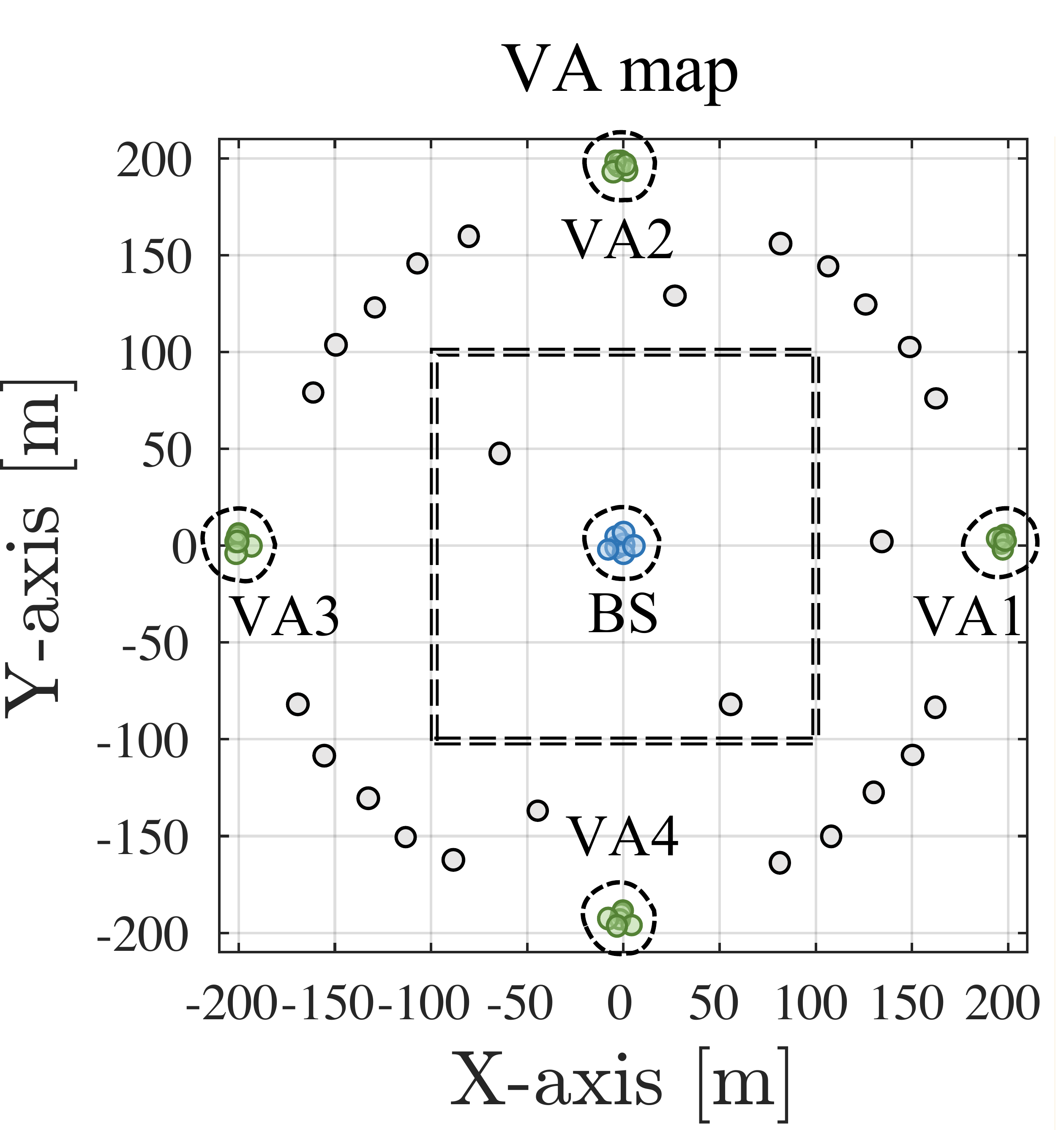}}
\subfigure[\label{Fig:birthc}]{\includegraphics[width=0.45\linewidth]{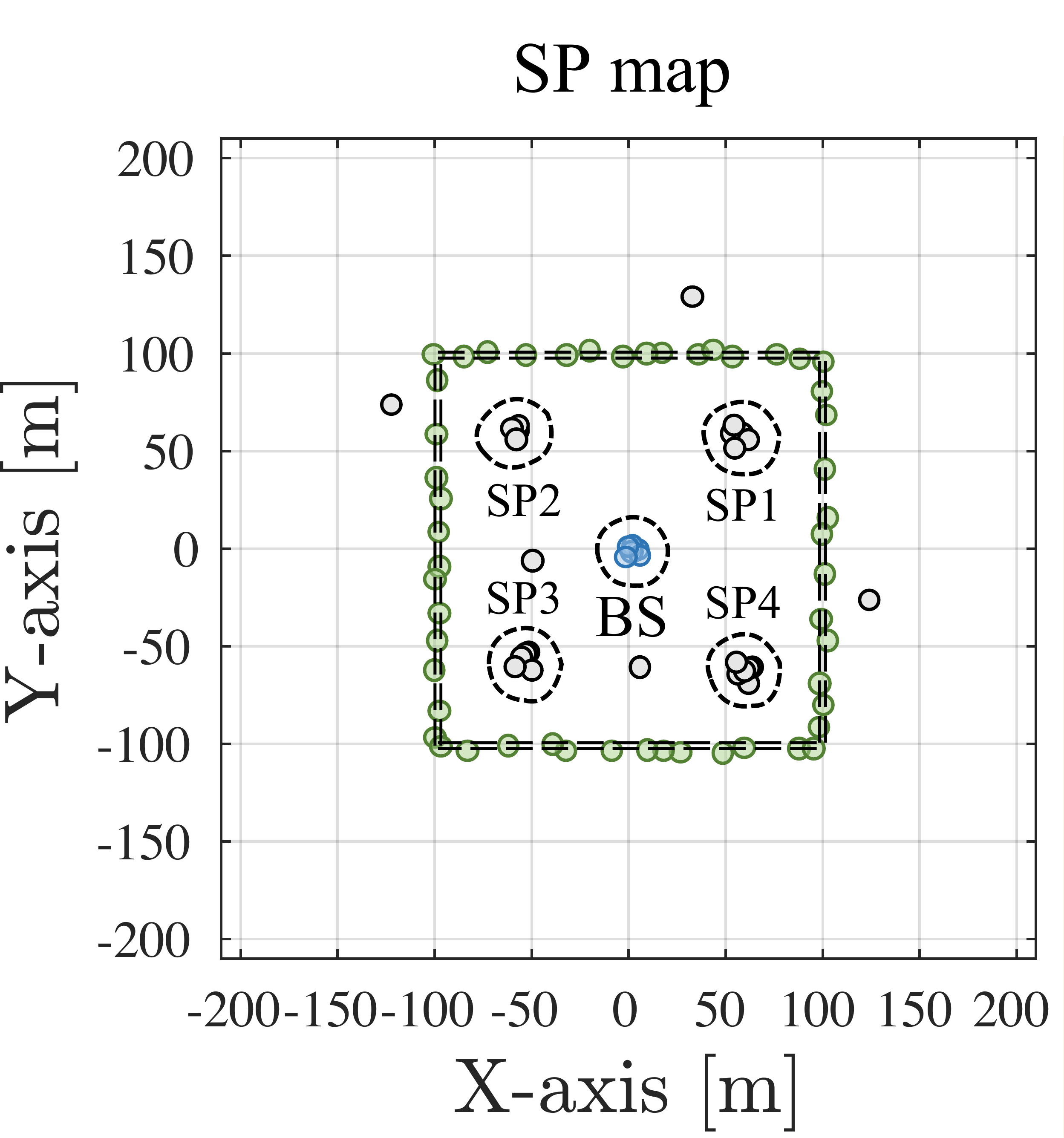}}
\subfigure[\label{Fig:birthd}]{\includegraphics[width=0.45\linewidth]{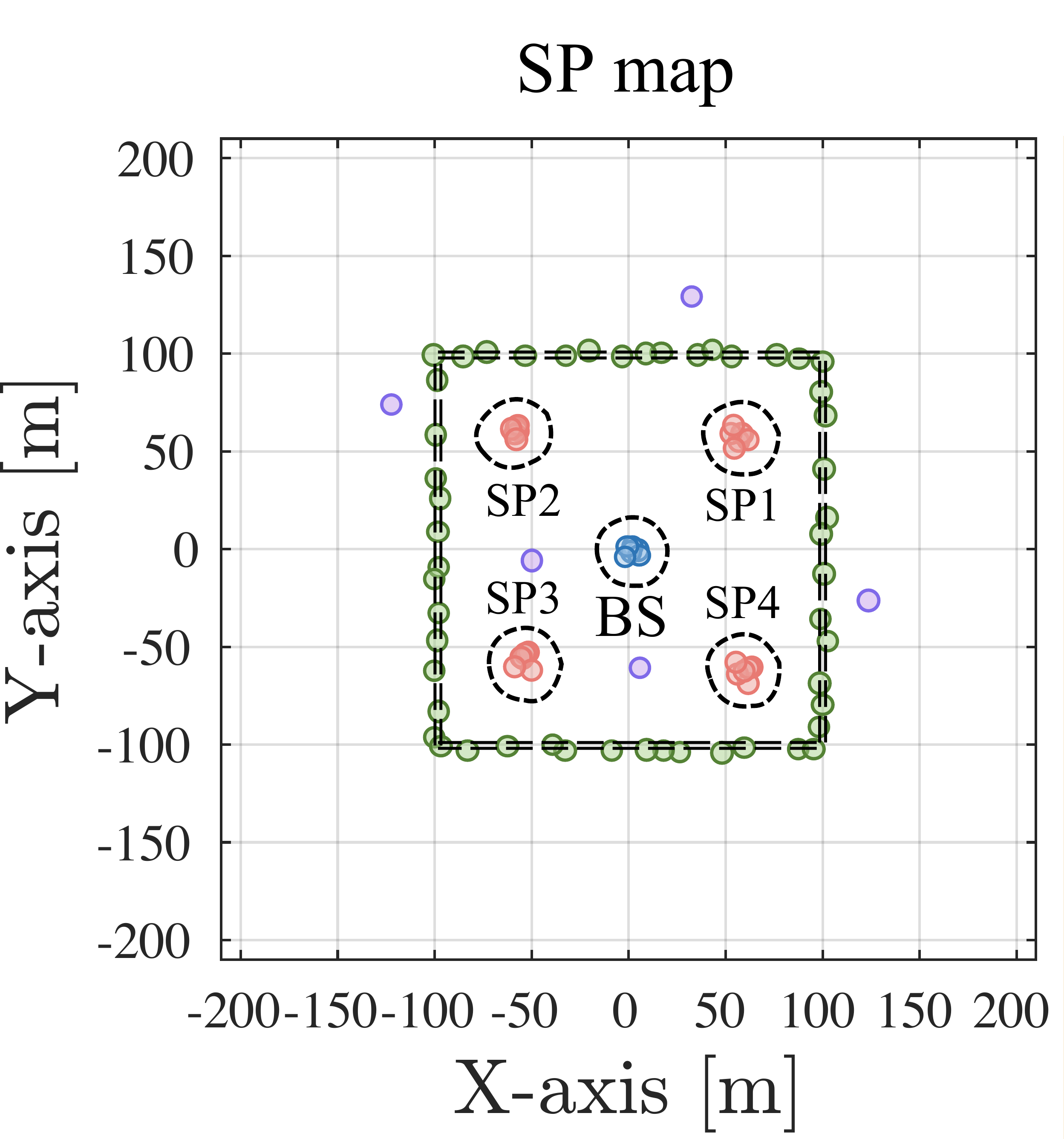}}
\caption{Example of VA/SP birth point using measurements (1 BS, 4 VA, 4 SP), (a) VA birth generation using measurements $\mathbf{z}_k$, (b) After VA mapping, we can identify VA but SP and clutter, (c) SP birth generation using $\mathbf{z}_k$, and we can recognize what VA measurements are by VA mapping, (d) Finally, we can identify SP using SP mapping.}\label{Fig:birth}
\end{figure} 


Fig. \ref{Fig:birth} shows an example of mapping using DP clustering in a radio wave environment consisting of 1 BS, 1 vehicle, 4 VAs due to 4 walls, and 4 SPs.
We go through the work of making the measurements into VA/SP birth points generated by the received signals for mapping.
The VA birth point is the VA position, assuming that the signal was received from the VA (i.e., reflected by the wall), and so is the SP birth point.
We call the map that shows the birth points of the measurements the birth map.

VA and SP measurements represent the VAs and SPs positions in the VA map and SP map.
For example, in Fig. \ref{Fig:birtha} to Fig. \ref{Fig:birthb}, the VA measurements form the VA birth points around VA1$\sim$VA4.
On the other hand, the birth points of the SP measurements and the clutters do not concentrate on one point.
Note that the LOS measurements indicate the position of BS in the VA map, and this makes it possible to identify LOS path measurement $\mathbf{z}^{\text{LOS}}_k$ from $\mathcal{Z}_k$.

As shown in Fig. \ref{Fig:birtha} to Fig. \ref{Fig:birthd}, birth points are intensively formed at the location of objects necessary for mapping, and we can cluster the intensively formed birth points into one group using DP in each birth map. Mapping using DP follows the following sequence. 
First, a VA map is generated, as shown in Fig. \ref{Fig:birtha}, and the birth points of the VA and LOS measurements of the VA map represent the location of VA and BS, respectively. These birth points representing the object can be clustered into a group through DP, and the birth point of the SP measurements and the clutter do not form a cluster in the DP clustering process. Through this, VA mapping is possible, as shown in Fig. \ref{Fig:birthb}, and the VA and LOS measurements are colored to express the distinction.
In the next step, we form the SP map as Fig. \ref{Fig:birthc}. We already know the VA and LOS measurements so that we can exclude them from DP clustering.
Likewise, the birth point of clutter is not clustered in the SP map, so SP mapping is possible, and finally, the result of Fig. \ref{Fig:birth} (d) is obtained.

\section{Performance Evaluation}
\subsection{Simulation Environment}
\begin{table}[t]
\centering
\caption{Simulation parameter units}
\label{tab:unit}
\begin{tabular}{ll}
\hlineB{3}
Parameter & Units \\ \hline \hline
$\text{diag}(\mathbf{Q})$ 
& $[\text{m}^2, \text{m}^2, \text{m}^2, \text{rad}^2, \text{rad}^2, \text{rad}^2, \text{rad}^2]$ \\
$\mathbf{s}_0, \sigma_0$ 
& $[\text{m}, \text{m}, \text{m},  \text{rad}, \text{m/s}, \text{rad/s}, \text{m}]^\top$\\
$\text{diag}(\mathbf{R})$ 
& $[\text{m}^2, \text{rad}^2, \text{rad}^2, \text{rad}^2 , \text{rad}^2]$\\
$\mathrm{diag}(\bm{\Sigma}), \mathrm{diag}(\bm{\Sigma}_0)$ 
& $[\text{m}, \text{m}]$\\
$\bm{\mu}_0$ 
& $[\text{m}, \text{m}, \text{m}]$\\
\hlineB{3}
\end{tabular}
\end{table}
We consider a vehicle that moves along a circular road for $K_{\text{max}} = 40$ with an interval of 0.5 seconds. 
The MATLAB simulation was conducted using the parameters as follows. 
We set $\mathbf{Q}$, the covariance noise matrix as $\mathrm{diag}[\sigma^2_x, \sigma^2_y, 0, \sigma^2_\alpha, 0, 0, \sigma^2_B]$, and $\sigma^2_x = 0.2$, $\sigma^2_y = 0.2$, $\sigma^2_\alpha = 0.01$, $\sigma^2_B = 0.2$. 
The initial state of the vehicle was set as $\mathbf{s}_0 = [0.7285, 0, 0, \pi/2, 22.22, \pi/10, 300]^\top$ and assume that the translation speed and turn-rate are known. 
The initial prior of the vehicle state follows Gaussian distribution, the standard deviation $\sigma_0$ is set to $[0.3, 0.3, 0, 0.3, 0, 0, 0.3]^\top$, and $\mathbf{R}$ is set to $\mathrm{diag}[10^{-2}, 10^{-4}, 10^{-4}, 10^{-4} ,10^{-4}]$.
For DP, we set $\omega$, $\bm{\mu}_0$, and $\bm{\Sigma}^0$ as 0.9, $[0,0,0]^\top$, and $\text{diag}[100, 100, 100]$, respectively and the units of simulation parameters are listed in Table \ref{tab:unit}.


The BS is located at $[0,0,40]^\top$, and four VAs are located at $[200,0,40]^\top,$ $[0,200,40]^\top,$ $[0,200,40]^\top,$ $[0,-200,40]^\top$, with unit $\text{m}$.
Four SPs are located at $[65,65,z_\mathrm{SP}]^\top,$ $[-65,65,z_\mathrm{SP}]^\top,$ $[-65,-65,z_\mathrm{SP}]^\top,$ $[65,-65,z_\mathrm{SP}]^\top$, with unit $\text{m}$ and $z_\text{SP} \sim \mathcal{U}(0,40).$ 
We set the detection probability $p_\text{D} = 0.9 $ within the field of view (FoV), the SP FoV is 50 m and VAs are always visible. We consider clutter intensity $c(\mathbf{z})$ follows Poisson distribution as $ \lambda /(4R_\text{max}\pi^4)$ as the average of the number of clutter measurements $\lambda =1 $, and the maximum sensing range $R_\text{max} = 200$ m. We use the average of the generalized optimal subpattern assignment (GOSPA) distance \cite{GOSPA} for measuring the mapping performance, and the parameter settings as \cite{HWTWC} for calculating GOSPA distance was used. Simulation results were obtained by averaging over 500 Monte Carlo runs. 

\subsection{Simulation Results}

\begin{figure}[t]
\begin{centering}
\includegraphics[width=1\columnwidth]{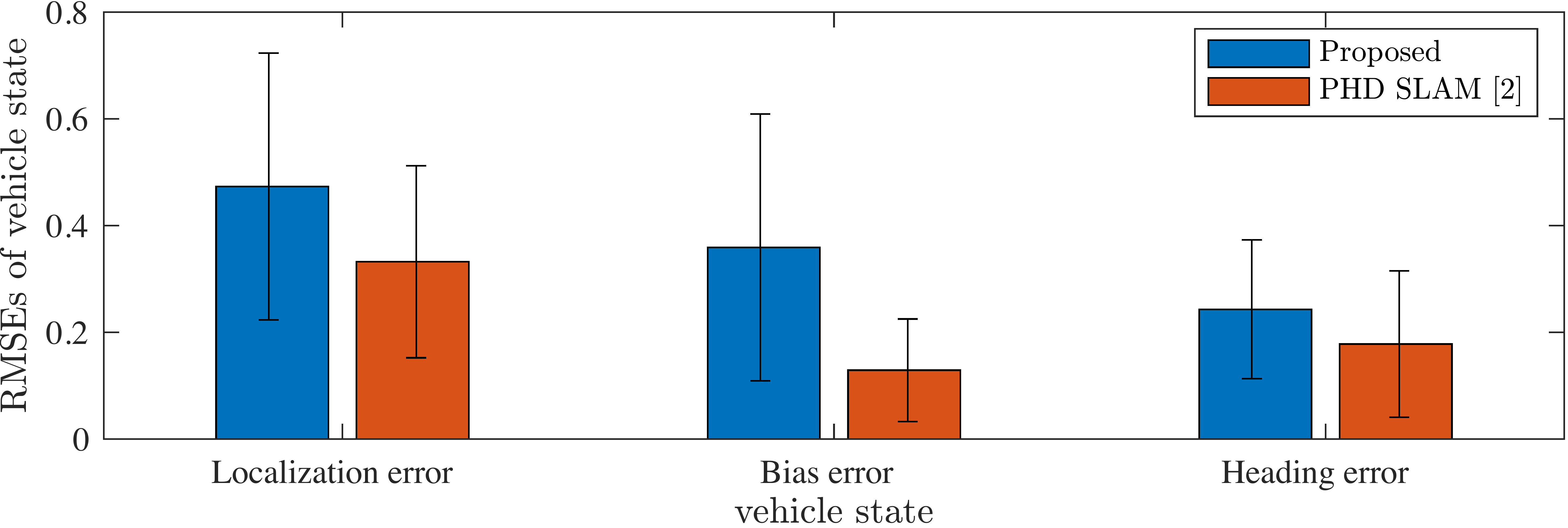}
\caption{MAE and RMSEs of vehicle state estimates (vehicle location, clock bias, and heading) by the proposed method compared to \cite{HWTWC}.} \label{Fig:Vehicle_state}
\end{centering}
\end{figure}


\begin{figure}[t]
\begin{centering}
    \subfigure[GOSPA_VA][\label{GOSPA_VA}]
{\includegraphics[width=1\columnwidth]{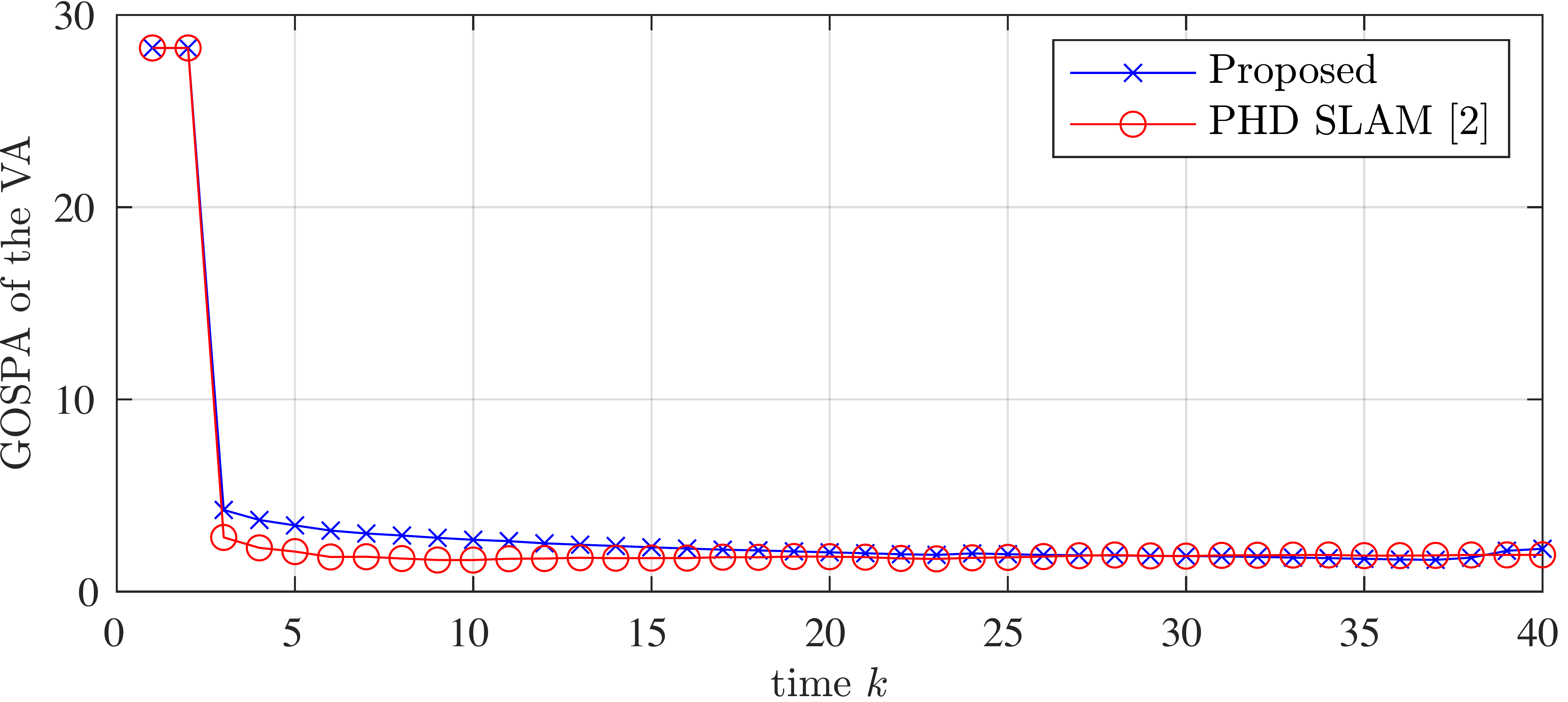}}\hfill 
\subfigure[GOSPA_SP][\label{GOSPA_SP}]
{\includegraphics[width=1\columnwidth]{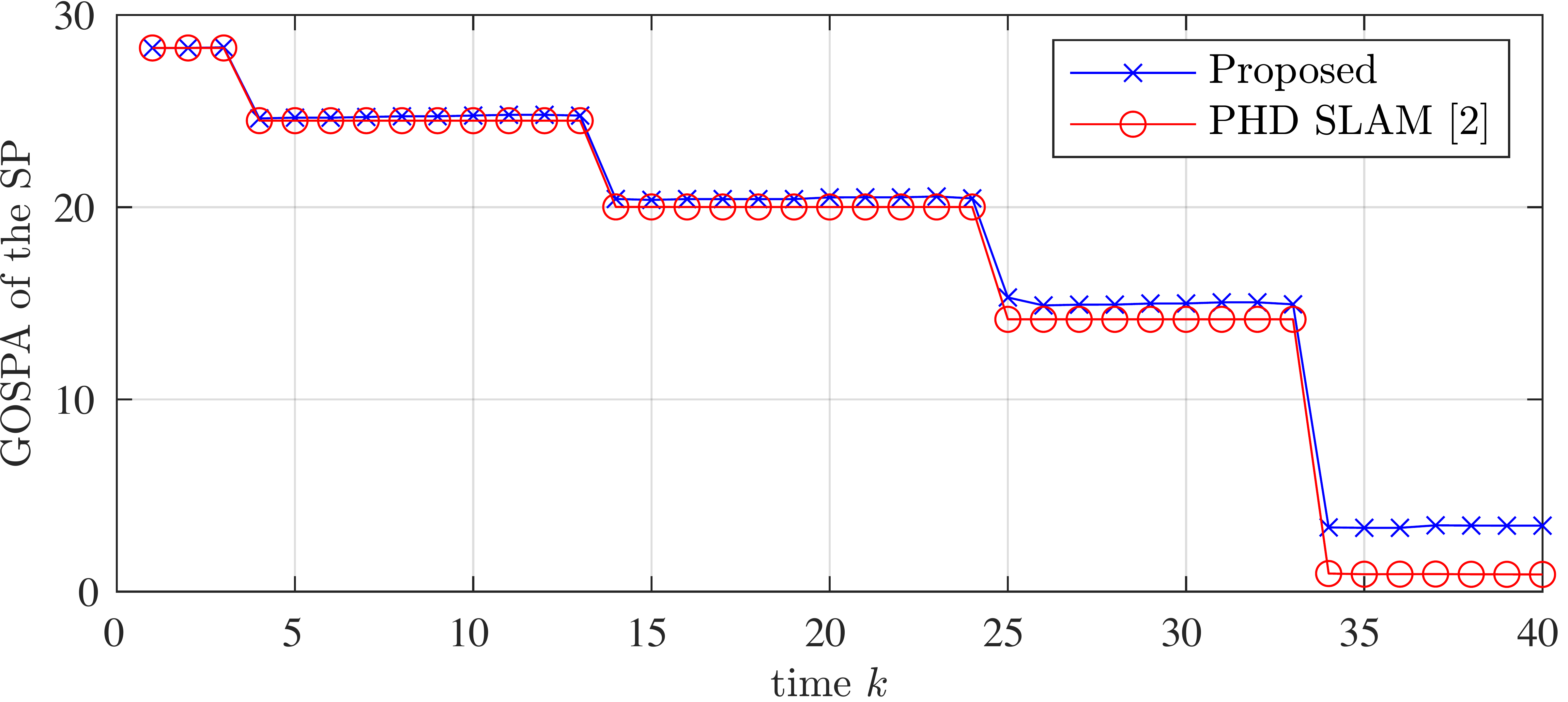}}
\hfill
\caption{Average GOSPA of (a) VA and (b) SP by the proposed method compared to \cite{HWTWC}.}\label{Fig:GOSPA}
\end{centering}
\end{figure}

To evaluate the performance of the proposed algorithm, we analyzed the vehicle position, clock bias, and vehicle heading of the vehicle state by mean absolute error (MAE), root means square error (RMSE), and the VA and SP positions using GOSPA.
\subsubsection{Localization}
Fig.~\ref{Fig:Vehicle_state} shows the MAE for estimated vehicle’s location and RMSEs for estimated clock bias and heading of the proposed DP SLAM compared to SLAM using Rao-Blackwellized PHD filter\cite{HWTWC} which the number of particles, $N_p$ is 2000. 
By comparing the results, estimation of the vehicle state through the proposed method has a slight performance drop, but we confirmed that there is a distinct gain of the complexity.
The average running time consumed by the proposed algorithm is 3.5 seconds, whereas the case of \cite{HWTWC} is more than 8000 seconds per 1 Monte Carlo trial.

\subsubsection{Mapping}
Fig.~\ref{Fig:GOSPA} represents the mapping performance of the proposed DP SLAM compared to \cite{HWTWC}.
In the case of the VA, Fig. \ref{GOSPA_VA} shows the average GOSPA of the VAs for the proposed method. 
The GOSPA of the proposed method is higher than that of \cite{HWTWC} at the beginning, but both methods show similar VA estimation performance over time.
Fig. \ref{GOSPA_SP} shows the SPs' average GOSPA of both methods.
Because of the limited FoV of the vehicle, the SP is detected only at a specific time to the vehicle, and this shows that the SP GOSPA stepwise decreases with time. 
In the end, from $k$ = 34, all SPs were detected, and GOSPA was finally reduced.
Compared with \cite{HWTWC}, as the detected SP increases, the SP estimation error is accumulated, and the difference in GOSPA increases gradually.
We confirmed the trade-off between computational complexity in mapping performance as state estimation performance.

\section{Conclusions}
In this paper, we proposed a DP-based SLAM for vehicle localization and mapping in-vehicle networks using 5G mmWave communication links.
At each time, the birth distributions, driven by the measurements, will be assigned to each object through DP, and environment mapping is obtained.
We confirmed that DP SLAM has the complexity gain while sustaining the SLAM accuracy compared to the Rao-Blackwellized PHD-SLAM filter.


\section*{Acknowledgment}
\scriptsize{This research was supported by the MSIT (Ministry of Science and ICT), Korea, under the ITRC (Information Technology Research Center) support program (IITP-2021-2017-0-01637) supervised by the IITP (Institute for Information \& Communications Technology Planning \& Evaluation).}

\bibliographystyle{IEEEtran}

\bibliography{bibliography.bbl}

\end{document}